\documentclass[english,12pt,aps,prd,a4paper,preprintnumbers,floatfix,nofootinbib,showpacs,superscriptaddress, notitlepage]{revtex4-1}
\pdfoutput=1
\usepackage{graphicx}
\usepackage{rotate}
\usepackage[utf8]{inputenc}

\usepackage{amsmath}
\usepackage{amssymb}
\usepackage{float}
\usepackage{comment}
\usepackage{soul}
\usepackage[usenames,dvipsnames]{color}
\usepackage[colorinlistoftodos]{todonotes}
\usepackage{amsmath}
\usepackage{amssymb}
\usepackage[colorlinks=true,citecolor=blue,urlcolor=blue, pdfborder={0 0 0}]{hyperref}
\usepackage[normalem]{ulem}
\newcommand {\ignore}[1]{}

\usepackage{makecell}
\definecolor{darkred}{rgb}{0.6,0,0}

\usepackage{gensymb}
\newcommand{\bl}[1]{{\color{blue}  #1}}

\usepackage[english]{babel}
\usepackage{blindtext}
\usepackage{tabularx}
\usepackage{subfig}

\begin{document}

\title{mRNA translation from a unidirectional traffic perspective}

\author{Binil Shyam T V}
\affiliation{Department of Biological Sciences, Indian Institute of Science Education and Research (IISER), Bhopal - 462066, Madhya Pradesh, India.}
\author{Rati Sharma}\email{rati@iiserb.ac.in}
\affiliation{Department of Chemistry, Indian Institute of Science Education and Research (IISER), Bhopal - 462066, Madhya Pradesh, India.}

\begin{abstract}
\noindent

    mRNA translation is a crucial process that leads to protein synthesis in living cells. Therefore, it is a process that needs to work optimally for a cell to stay healthy and alive. With advancements in microscopy and novel experimental techniques, a lot of the intricate details about the translation mechanism are now known. However, the why and how of this mechanism are still ill understood, and therefore, is an active area of research. Theoretical studies of mRNA translation typically view it in terms of the Totally Asymmetric Simple Exclusion Process or TASEP. Various works have used the TASEP model in order to study a wide range of phenomena and factors affecting translation, such as ribosome traffic on an mRNA under noisy (codon-dependent or otherwise) conditions, ribosome stalling, premature termination, ribosome reinitiation and dropoff, codon-dependent elongation and competition among mRNA for ribosomes, among others. In this review, we relay the history and physics of the translation process in terms of the TASEP framework. In particular, we discuss the viability and evolution of this model and its limitations while also formulating the reasons behind its success. Finally, we also identify gaps in the existing literature and suggest possible extensions and applications that will lead to a better understanding of ribosome traffic on the mRNA.

\end{abstract}

\maketitle

\section{Introduction}
\noindent
Translation of mRNA into protein is one of the fundamental processes in the central dogma of life that describes the flow of information within cells. The central dogma states that the genetic information encoded within the DNA gets transcribed into RNA which then gets translated to a functional form as a protein \cite{alberts2002molecular, cooper2007cell}. The translation process, therefore, is an important cog in the wheel that scientists have been trying to elucidate in detail ever since the structure of DNA was discovered in the 1950s. 
\\~\\
Several important advancements in experimental techniques such as ribosome profiling and single molecule imaging have led to a more detailed understanding of the translation process \cite{gobet2017ribosome, calviello2017beyond, morisaki2018quantifying, sonneveld2020heterogeneity, neelagandan2020determines}. In tandem with these experimental studies, theoretical studies of the translation mechanism have also progressed and led to successful explanations of the observed phenomena. mRNA translation is a process where a ribosome assembly binds to the 5' end of the mRNA and moves unidirectionally from codon to codon to the 3' end while also in the process adding amino acids to the growing oligomeric chain. The assembly exits the mRNA and simultaneously releases the complete protein once it reaches the 3' end. One of the successful theoretical modeling frameworks that neatly captures this ribosomal translocation and therefore has gained popularity in recent decades is the Totally Asymmetric Simple Exclusion Process (TASEP) framework. This framework views the translation process as a unidirectional traffic of ribosomes from one end to the other. Given its success in explaining several properties, such as, ribosome traffic, density and current \cite{macdonald1968kinetics, shaw2003totally, chou2011non}, it is pertinent to understand and ask the question as to what makes this framework so suitable for modeling translation.  
\\~\\
%
With this aim, in this review, we look at mRNA translation from the point of view of unidirectional traffic and understanding the features of the TASEP modeling framework. The kinds of problems addressed through the TASEP framework are vast ranging from protein synthesis and translation \cite{macdonald1968kinetics, shaw2003totally} to transcription \cite{Klumpp2008, Klumpp2009, Wang2014}, transport of cargo via molecular motors \cite{chou2011non, Neri2013, ciandrini2014stepping}, 
fungal growth \cite{Sugden2007, evans2007exclusion, sugden2007dynamically} and vehicular traffic \cite{Chowdhury2000, appert2011intersection}. Therefore, we also provide a history of the model, its role in explaining various phenomena and its application to theoretical studies of translation and other intracellular processes in various model organisms. 

\section{A primer on the translation mechanism}
\noindent
Before going into the specifics of the TASEP framework, let us first understand what translation is at the molecular level. The central dogma, essentially, comprises of two fundamental processes, transcription and translation \cite{alberts2002molecular, cooper2007cell}. The transcription process ensures that the information encoded in segments of DNA gets transferred into RNAs. RNAs, in turn, act as information carriers and regulators for protein synthesis and this process of producing proteins from RNAs is known as translation. However, there are a few major differences between DNA and RNA, the key players of the central dogma. DNAs are double stranded, whereas RNAs are single stranded. The DNA backbone is made up of deoxyribose sugars while the RNA backbone is composed of ribose sugars. Further, one of the four bases is different for RNA - Uracil instead of Thymine - in comparison to DNA. DNA can essentially be understood in terms of an information storage system, while RNA as an information carrier. RNA transcripts are therefore produced in large numbers as identical copies.
\\~\\
Different segments of DNA are transcribed into several different categories of RNA \cite{alberts2002molecular}. The major category is that of the messenger RNA (mRNA) that is copied from genes (segments that code for proteins) in the DNA. Other categories are ribosomal RNAs (rRNAs) that catalyze protein synthesis, transfer RNAs (tRNAs) that act as intermediaries between mRNA and amino acids, small nuclear RNAs (snRNAs) that regulate nuclear processes, small nucleolar RNAs (snoRNAs) that process or modify rRNAs, microRNAs (miRNAs) that regulate gene expression by inhibiting the function of specific mRNAs, small interfering RNAs (siRNAs) that cause gene silencing through degradation of mRNAs, piwi-interacting RNAs (piRNAs) that regulate piwi proteins and finally long noncoding RNAs (lncRNAs) that regulate various cellular processes. The translation process itself is carried out by mRNA with help from tRNAs and ribosomes.
\\~\\
Depending on the length of the gene, mRNAs are composed of a few hundred to a few thousand codons (a chain of three adjacent nucleotides). For protein synthesis to happen, these codons need to be read individually in succession so that the corresponding amino acids can be added to the growing polypeptide chain. There are 64 possible codons out of which 61 code for the 20 common amino acids and the remaining 3 primarily act as stop codons but, are also known to incorporate selenocysteine and pyrrolysine in some organisms \cite{zhang2005pyrrolysine}. Therefore, each amino acid corresponds to multiple codons, called synonymous codons. Further, apart from mRNAs themselves, the process of translation that eventually leads to protein synthesis involves two other key players, viz. ribosomes and tRNAs.  tRNAs act as adaptors that mediate recruitment and pairing of amino acids to codons. Specifically, one end of the tRNA binds to the amino acid, wherein the binding is mediated by an enzyme group called aminoacyl tRNA synthetases. The anticodon loop in the tRNA containing the complementary bases for base pairing with the particular codon of the mRNA is capable of forming a codon-anticodon pair. 
\\~\\
Ribosomes enter the picture following the binding of the tRNA with the amino acid. Ribosomes consist of two subuints where each subunit is composed of both rRNAs and proteins. For example, the 70S ribosome in prokaryotes consists of a smaller 30S subunit and a larger 50S subunit. Similarly, the 80S ribosome in eukaryotic cells is composed of the smaller 40S subunit and the larger 60S subunit. The entire process of translation is divided into three major steps, viz. initiation, elongation and termination with ribosomes playing an important role in all the three steps.
\\~\\
During the initiation process, a group of initiation factors first bind to the smaller subunit of the ribosome to form a complex. The mRNA and the initiator aminoacyl tRNA complex are then brought together to bind to the complex. The recognition of the initiator tRNA by the ribosomal complex then initiates the association of the larger ribosomal subunit to the complex. The formation of the complete ribosomal initiation complex at the start codon of the mRNA at the 5' end then triggers the peptide bond formation during the elongation process. 
\\~\\
Ribosomes possess three tRNA binding sites, called the E (exit), P (peptidyl) and the A (aminoacyl) sites. The initiator aminoacyl tRNA complex first binds to the P site. The next aminoacyl tRNA complex attaches itself to the A site. Following this, there is a peptide bond formation between the two amino acids on the P and A sites which is facilitated by the larger ribosomal subunit. This is followed by the transfer of the initiator amino acid to the A site as well, leaving the P site with just the tRNA. The complete ribosomal machinery then moves to the next codon leaving the A site empty and translocating the peptidyl tRNA to the P site and the uncharged tRNA to the E site. Finally, the association of a new aminoacyl tRNA to the now empty A site triggers the exit of the uncharged tRNA from the E site. 
\\~\\
This elongation process continues until the stop codon is reached at the A site. Release factors then recognize this signal and release the tRNA and the complete protein at this step and also trigger the dissociation of the ribosomal machinery from the mRNA. This process is called termination. In general, a new scanning ribosome can detect the start codon and initiate a new ribosome onto the transcript whenever the start codon is unoccupied and binding of a new ribosome is not physically hindered. There is a steady stream of ribosomes being initiated once a ribosome moves away from the start site during elongation, therefore a single mRNA is simultaneously translated by several ribosomes. This process continues until the mRNA degrades away. 
\\~\\
\begin{figure}[H]
    \centering
    \includegraphics[width=155mm]{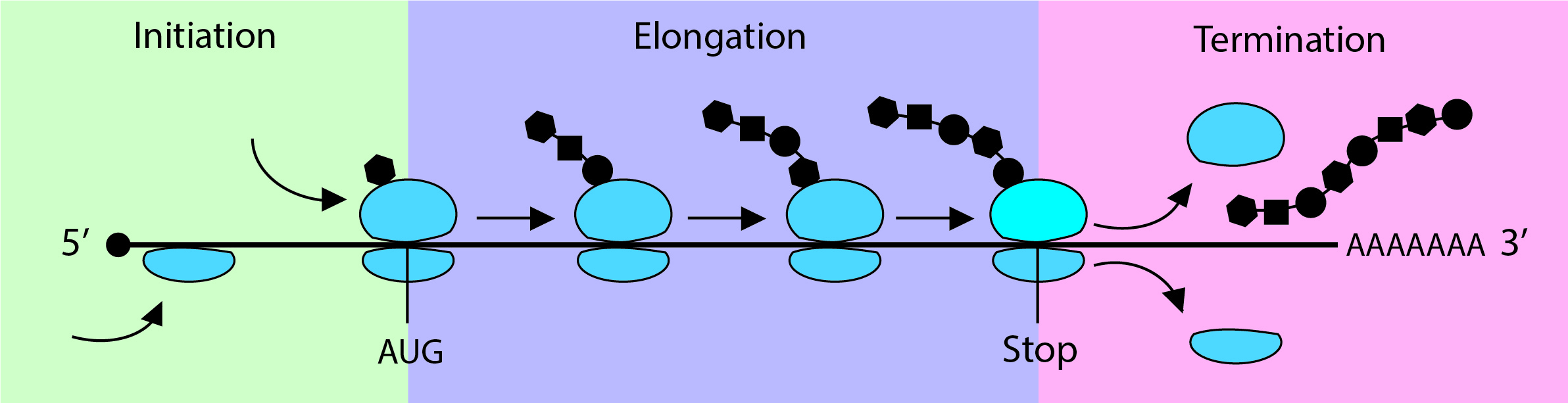}
    \caption{A representation of the different phases of translation [Adapted from \cite{cooper2007cell, sonneveld2020heterogeneity}]. \((1)\) Initiation, ribosomal subunits and initiator methionine tRNA along with several other translation initiation factors are recruited to the 5' UTR of mRNA. \((2)\) Elongation, ribosome and associated factros moves along the mRNA adding aminoacids corresponding to each codon. \((3)\) Termination, when ribosome reaches a stop codon, the polypeptide and ribosome are released from the mRNA}
    \label{fig:translation}
\end{figure}
\noindent
In a simplified picture of this entire process of translation \cite{macdonald1968kinetics, shaw2003totally} (as represented in Fig. \ref{fig:translation}), one generally considers the three steps in the following manner. The initiation step is determined by the rate of binding of the ribosomal machinery to the mRNA at the start site. In the elongation step, this ribosomal machinery moves from codon to codon on the mRNA at a certain rate and in the process successively adds amino acids to the polypeptide chain. Finally, when the ribosome reaches the stop site at the 3' end of the mRNA, the process terminates with a certain rate and releases the protein, mRNA, and the ribosome. 
\\~\\
The following sections will delve deeper into the physics of the entire process through the TASEP framework.  

\section{The TASEP Framework}
\begin{figure}[H]
    \centering
    \includegraphics[width=130mm]{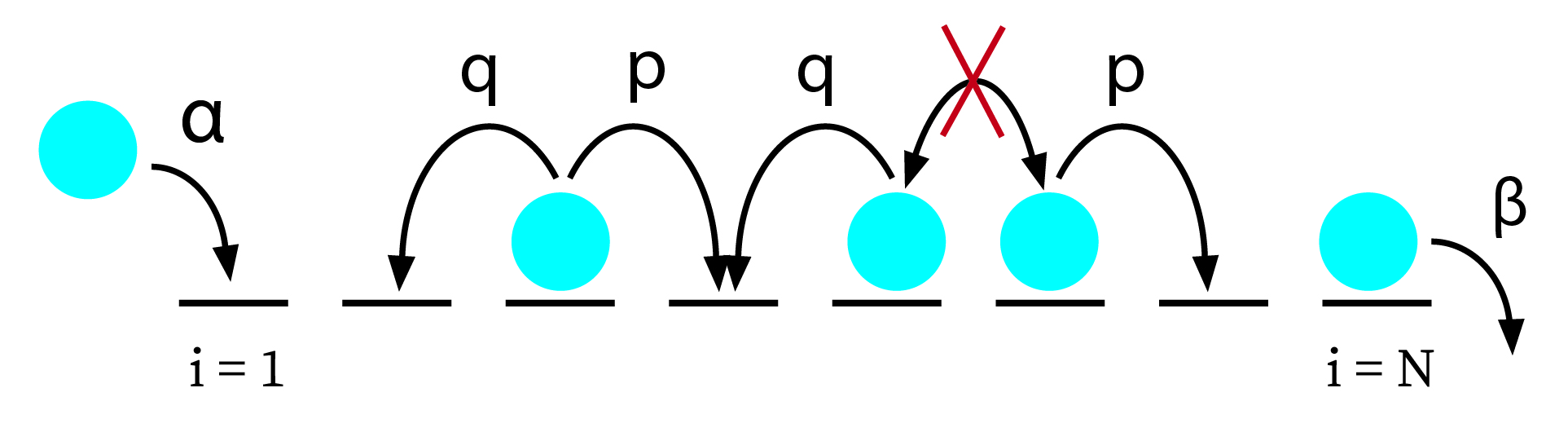}
    \caption{A representation of Asymmetric Simple Exclusion Process (\textbf{ASEP}) model. Particles enters the lattice of N sites at site \(i = 1\) with a rate \(\alpha\) when it is vacant. Particles exit the lattice at site \(i = N\) with a rate \(\beta\). In this general ASEP framework the particles can hop bidirectionally with a rate \(p\) and \(q\) while strictly following the exclusion rule. If the neighbouring site is occupied the particle retains its original position while waiting for the neighbouring site to become vacant.}
    \label{fig:asep}
\end{figure}

\begin{figure}[H]
    \centering
    \includegraphics[width=70mm]{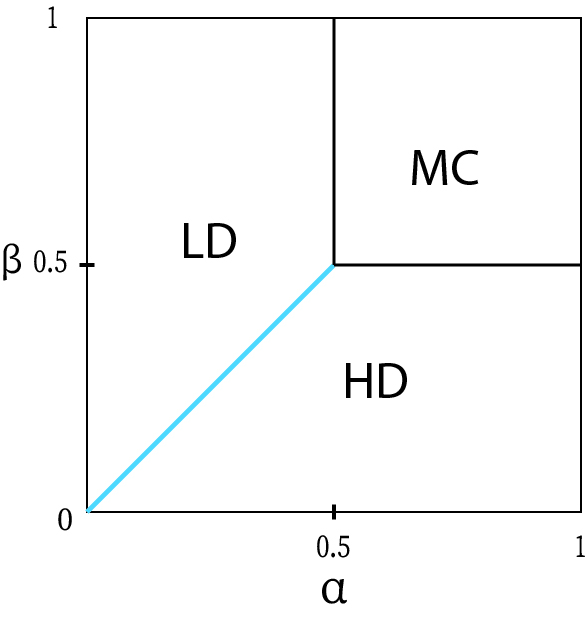}
    \caption{TASEP phase diagram, System when \(\alpha>\beta\;,\; \beta<1/2\) is in high density (HD) phase, which causes accumulation of particles on the lattice. The system with \(\alpha<\beta\;,\; \alpha<1/2\) is in the low density (LD) phase which causes low occupancy on the lattice and when \(\alpha,\;\beta>1/2\), the system is in the maximal current (MC) phase. The transition line with \(\alpha=\beta < 1/2\) (represented in blue) is where a superposition of states is seen and across it the transition is discontinuous.}
    \label{fig:phase}
\end{figure}

\noindent
The first consideration before developing a model for any process is understanding whether the system under consideration is in equilibrium or not. For a system in equilibrium, one can determine its distribution by first formulating and re-writing it in terms of an ensemble of microstates, all equally probable. However, if the system is out of equilibrium, a complete picture can only be obtained through a probabilistic interpretation and evolution of its states \cite{chou2011non}.
Specifically, a system of particles coupled at two ends by a particle reservoir with different chemical potentials, after some initial transients, reaches a state of constant non zero mean flux, placing the system in a non equilibrium steady state (NESS). The most distinguishing feature of NESS is the irreversible exchange of some physical quantity between the system and the surroundings \cite{blythe2007nonequilibrium}, leading to a non zero current (of energy, mass, or other quantities) and a nonzero chemical potential gradient in the system \cite{qian2006open}. Recognizing the  importance and impact of non equilibrium physics in various mathematical and biological systems, several approaches have been taken to understand processes far from equilibrium \cite{bertini2007stochastic}. The ideal approach to this study would be to model particle interactions at the atomic level. However, this is often non-trivial and therefore, the more realistic approach is to adopt a simplified model that captures the essence of the system.  
\\~\\
\textbf{Driven lattice gas model.} The simplest framework that can capture the non-zero flux across a system is the driven lattice gas model \cite{van1985excess}. This model system is composed of a D dimensional lattice with L sites, where each site can accommodate at most one particle. The particles jump stochastically to neighbouring sites according to rates that depend on the local environment and an external field that biases the jump in some direction \cite{krug1991boundary}. This same approach was used to develop the Ising model of ferromagnetism in order to explain how interaction between neighbouring molecules in a crystal can give rise to correlated behaviour and phase transition \cite{majewski2001ising}.
\\~\\
\textbf{Asymmetric Simple Exclusion Process} (\textbf{ASEP}) is the simplest example of a driven diffusive system. The ASEP model, represented in Fig. \ref{fig:asep}, describes conformations of point like particles on a one dimensional lattice comprising of L sites, with the state of each site represented by \(x_{i}\). In the case of an open system 
then, there are three possible events that can take place on the lattice. These are (i) A new particle can enter the lattice at site $i = 1$ with rate $\alpha$, (ii) A particle can exit the lattice at site $i = L$ with a rate $\beta$ and (iii) each particle on the lattice can jump to its vacant neighbouring site either on the left or the right with a rate $q$ and $p$, respectively. 
During each time step $dt$ (i.e. the infinitesimally small time interval when a reaction is likely to occur), any one of these events may occur with some probability (as a function of their propensity). Each successive event is independent of the previous one imparting stochasticity to the model. At a given point in time, the system is therefore in one of its \(2^L\) possible configurations and evolves as a Markov process \cite{lazarescu2011exact}. A schematic representation of this process is presented in Fig. \ref{fig:asep}. Several exact results for computing different properties of ASEP have been derived either using the matrix product ansatz \cite{blythe2007nonequilibrium} or the Bethe ansatz \cite{derrida1998exactly}.
\\~\\
\noindent
Within the general ASEP model if \(q = p\) , the system is referred to as the symmetric simple exclusion process (SSEP). On the other hand, if \(q = 0\) the system becomes the classical Totally asymmetric simple exclusion process (TASEP) and if \(0<q<p\) the system is known as the Partially asymmetric simple exclusion process (PASEP). Since these models allow for a simplification of complex interactions, they have been applied to various fields in science \cite{chou2011non} including biological \cite{kuan2016motor, chou2011non} and non biological transport \cite{Chowdhury2000, appert2011intersection, hilhorst2012multi, kouhi2019interpreting},  random walk in a disordered environment \cite{huveneers2015random}, ion channel dynamics \cite{chou1999entropy} and protein synthesis \cite{zia2011modeling}, among others.
\\~\\
\textbf{Totally Asymmetric Simple Exclusion Process} (\textbf{TASEP}) models are one dimensional interacting particle systems that can incorporate asymmetric events on a lattice framework. TASEP follows the same rules as the ASEP except, here, the movement of particles is restricted to one direction, conventionally left to right. TASEP is mainly studied under two boundary conditions, open and periodic. Under the open boundary condition, particles enter the lattice one at a time from the left reservoir with a rate \(\alpha\) when the first lattice site (\(i=1\)) is vacant and exit the lattice from the last lattice site  (\(i=L\)) with a rate \(\beta\). These rates correspond to a coupling of the system to reservoirs of densities \(\alpha\) and \(1-\beta\), respectively \cite{kolomeisky1998phase}. In the bulk system, the particle at site \(i\) hops to site \(i+1\) with a rate \(p\) if the site is vacant. Under the periodic boundary condition, the particle again moves from site \(i\) to site \(i+1\) with a rate \(p\) but, when the particle reaches the last lattice site \(i=L\), it hops back to the first lattice site $(i=1)$ if it is vacant. Therefore, the number of particles are conserved. Unidirectional hopping of particles along the lattice and particle interactions following the exclusion rule that forbids particles from overlapping or overtaking are some of the characteristics of these frameworks. The TASEP system has been studied extensively and solved analytically for the case with open boundaries. Under these conditions and with the particle hopping rates equal to 1 (i.e., $p=1$), the lattice sites have been shown to have non-uniform densities \cite{derrida1992exact, schutz1993phase}.
\\~\\
In TASEP models with open boundaries, non equilibrium phase transitions can be induced by varying the two boundary parameters. A phase transition is observed in terms of a sudden change in the global ensemble average quantity such as the particle current or bulk density. Depending on the boundary conditions, the system can be in one of the three phases, i.e., High density (HD), Low density (LD) and Maximal current (MC) \cite{krug1991boundary, derrida1993exact, kolomeisky1998phase, popkov1999steady, blythe2007nonequilibrium, wood2020combinatorial}, also represented in Fig. \ref{fig:phase}. The HD phase refers to a high rate of particle injection (high \(\alpha\)) on to the lattice coupled with slower rate of ejection (low \(\beta\)) at the termination site which causes accumulation of particles on the lattice. The LD phase refers to a low rate of particle injection (low \(\alpha\)) and higher rate of particle ejection (high \(\beta\)) that eventually leads to low particle occupancy on the lattice. The MC phase refers to the system where particles enter and exit the lattice at a fast rate leading to high particle current through the system. The transition between the phases are characterised by sharp changes in the steady state particle current \(J\) and particle density $\rho$. In the general case of ASEP, where transition to the left is given by the rate $q$ and transition to the right is given by $p=1$, the phases depend on \(\alpha\), \(\beta\) and \(q\) in the following manner \cite{wood2020combinatorial}.
\begin{itemize}
    \item \textbf{HD}: if $\alpha>\beta\;,\; \beta<(1-q)/2$, then the current $J = \beta(1-q-\beta)/(1-q)$
    \item \textbf{LD}: if $\alpha<\beta\;,\; \alpha<(1-q)/2$, then the current $J = \alpha(1-q -\alpha)/(1-q)$
    \item \textbf{MC}: if $\alpha,\beta>(1-q)/2$, then the current $J = (1-q)/4$
\end{itemize}
In the case of the TASEP framework (i.e., when $q = 0$ and $p=1$) and when the bulk density of the system is also well defined, the current and density in the three phases depend on the rates as per the following.
\begin{itemize}
    \item \textbf{HD}: if \(\alpha > \beta\) and \(\beta < 1/2\), then the current \(J = \beta(1-\beta)\) and density \(\rho = 1-\beta\)
    \item \textbf{LD}: if \(\alpha<\beta\); and \(\alpha<1/2\) , then the current \(J = \alpha(1-\alpha)\) and density \(\rho = \alpha\)
    \item \textbf{MC}: if \(\alpha\), \(\beta>1/2\), then the current \(J = 1/4\) and density \(\rho = 1/2\)
\end{itemize}
Although the phase diagram in Fig. \ref{fig:phase} has been obtained by considering $\alpha<1$, $\beta<1$ and $p=1$, the same can also be obtained when $p\neq1$ by scaling the $\alpha$ and $\beta$ to $\alpha/p$ and $\beta/p$, respectively \cite{rajewsky1998asymmetric}.
\\~\\
In addition to the above three phases, a fourth phase exists when $\alpha = \beta$ and both $\alpha , \beta<1/2$. This phase, known as the shock phase (SP), leads to a state where both high density and low density phases coexist distinguishable by a domain wall which diffuses stochastically along the lattice and therefore cannot be localised \cite{kolomeisky1998phase}. Specifically, the line $\alpha=\beta < 1/2$ indicates a first order transition \cite{derrida1993exact}. A superposition of states exists along this first order transition line, wherein a shock can travel randomly between a region of density $\alpha$ and a region of density $1-\beta$. When the system exists near this transition line, small fluctuations can push it from one density phase to the other, causing a change in the bulk density that then leads to the formation of a shock traveling from the right to the left boundary at a linear speed \cite{elboim2022mixing}. The transitions between MC and the other two phases (HD and LD) are continuous, while the transitions across the HD\(-\)LD phases are discontinuous \cite{zia2011modeling, blythe2007nonequilibrium, kolomeisky1998phase}.
\\~\\
Even though the classical TASEP framework provides valuable insights into one dimensional many particle processes, modifications need to be carried out in order to explain more complex processes. The classical TASEP framework works with particles that occupy one lattice site and where jump rates are homogeneous across all sites.  However, a more generalised version of this framework, called the $\ell$-TASEP can incorporate extended particles, i.e., particles that occupy more than one lattice site and where jump rates could be inhomogeneous as well \cite{shaw2003totally,erdmann2020key}. This generalized $\ell$-TASEP framework has also been incorporated into a computational tool known as EGGTART for easy quantification and visualization of the ribosome dynamics and associated translation \cite{erdmann2021eggtart}. This tool is also successful in recovering the phase diagram, current and density in the hydrodynamic limit and is qualitatively similar to that obtained in the classical TASEP framework, where $\ell=1$. Further, the TASEP model in general, discretises a process which in nature is continuous, for the ease of carrying out calculations. However, as we shrink the lattice space or the scale of the model, it becomes a better representation of the originally continuous process and at the continuum limit we can recover the original process.
\\~\\
Another extension of the TASEP framework is the Ribosome flow model (RFM) \cite{reuveni2011genome}. It is a mean-field approximation of the $\ell$-TASEP model with open boundaries where the ribosome dynamics are studied in the steady state. Specifically, in the RFM framework, the flow of ribosomes are first modelled as a set of coupled differential equations for the probability of occupancy at each site. Then the dynamics are averaged over several transcripts to obtain a steady state description such that the occupation probabilities are constant in time. Further, the ribosome flow into and out of each site, modelled in terms of these occupation probabilities, are each equated to the steady state rate of protein production. It is also important to note that this rate corresponds directly to the steady state rate of the ribosome exiting an mRNA transcript. In setting up the framework and solving the set of equations, RFM makes two approximations. These are (i) coarse-graining the entire length of mRNA into chunks/groups of sites where each site has a normalized ribosome density and (ii) assuming a lack of correlations between any two sites. These approximations help in simplifying the model system while maintaining the core elements of the problem, which is, to model ribosome flow or dynamics along the mRNA. An improvement over the RFM framework is the Ribosome flow model with extended objects (RFMEO) where every ribosome on the mRNA lattice is explicitly assumed to cover $\ell$ sites with the site-dependent elongation rate considered only for the apriori assumed reader site among the $\ell$ sites \cite{zarai2017ribosome}. In comparison to the TASEP based models, the RFM framework can in general be treated more easily mathematically and analytically through the use of tools and techniques from the field of systems and control theory, which makes it the more attractive framework for some researchers. 
\\~\\
With the above presented background, we can now look into the applications of the TASEP framework to the process of translation.


\section{Understanding \lowercase{m}RNA translation dynamics through the TASEP framework}

\begin{figure}[H]
    \centering
    \includegraphics[width=\textwidth]{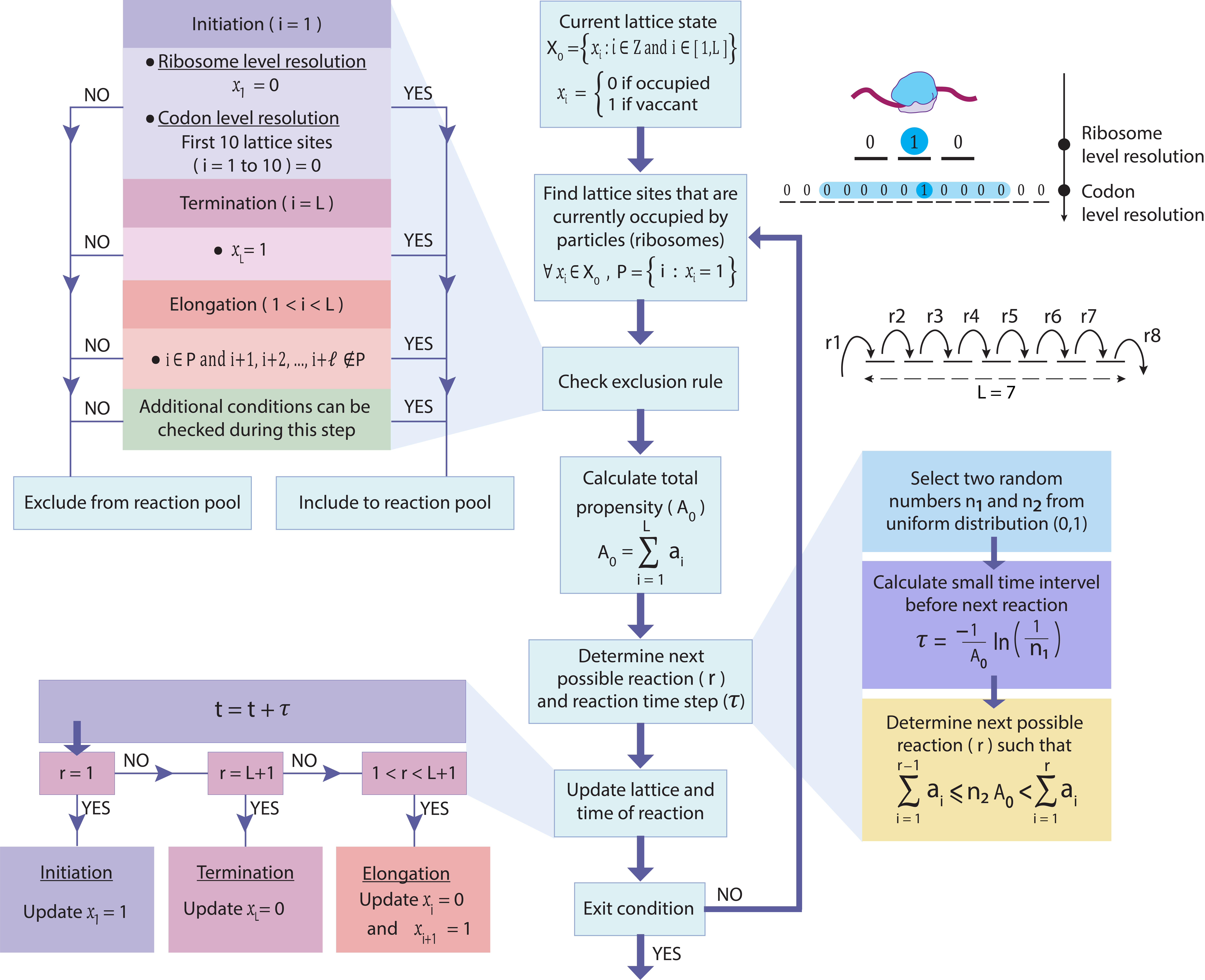}
    \caption{A flow chart representation of the TASEP model of translation with open boundaries and random sequential update following the Gillespie algorithm.}
    \label{fig:flowchart_tasep}
\end{figure}

\noindent
A successful implementation of a modeling framework to study any system depends on the identification of some of the key features of the process and modification of classical models to accommodate them. 
%
To this end, one of the main approaches to modeling protein synthesis in the 1960s was to use the deterministic route where the entire system could be set up mathematically in terms of ordinary differential equations (ODEs) \cite{von2012mathematical}. In this formalism, each codon-ribosome complex is considered as an independent chemical species and the forward movement of this complex on the lattice is viewed in terms of a chemical reaction. Therefore, a ribosome moving across an mRNA with $L$ lattice sites would be represented through $L-1$ ODEs. The number of variables in the set of ODEs would therefore be large and depend on the number of ribosomes as each new ribosome would comprise a new chemical species. This approach, in particular, was used to solve a set of ODEs that explained a system comprising of an mRNA lattice with 6 sites and 2 ribosomes \cite{gerst1965kinetics, garrick1967kinetics}.
\\~\\
However, these early ODE-based models became cumbersome and non-trivial to solve when applied to systems with increasing complexity. They could also not capture some of the characteristic features of mRNA translation, such as non overlapping ribosomes and single occupancy of a codon site \cite{zhao2014mrna}. A few decades later, as mentioned earlier, a mean field approach was used to again model the ribosome dynamics in terms of the ribosome occupancy probabilities through the RFM \cite{reuveni2011genome}. This deterministic approach, an improvement over the earlier ODE based models, captured codon-level kinetics of a ribosome on a full length mRNA and was used to predict translation rates, protein abundance and ribosome densities in the endogenous genes of \textit{E. coli}, \textit{S. cereviseae} and \textit{S. pombe}. It performed especially well when there was a significant correlation between protein abundance and the codon order on the transcript and therefore improved upon earlier predictions for cases where codon order was ignored but might have been an important consideration. RFM and its extensions have now been used to study many more biological problems in real systems, including maximizing the overall protein translation rate \cite{poker2014maximizing, zarai2016ribosomal}, initiation rate or elongation rate dependent sensitivity of mRNA translation to mutations \cite{poker2015sensitivity}, studying the competition for shared pool of ribosomes \cite{raveh2016model, Jain2022}, uncovering a link between optimal downregulation of mRNA translation and the elongation rate associated with the codon position \cite{zarai2017optimal}, studying networks of mRNAs \cite{nanikashvili2019networks}, ribosome dynamics along a ring \cite{raveh2015ribosome} and ribosome dropoff \cite{Jain2022}, among others. 
\\~\\
As an alternative to the \bl{early} deterministic models, statistical mechanics based approaches were \bl{also} explored to capture ribosome movement. The first set of statistical mechanics based models that was proposed considered probabilistic hopping of ribosomes on the mRNA lattice \cite{simha1963polymerization, zimmerman1965kinetics, macdonald1968kinetics, macdonald1969concerning}. In fact, protein synthesis can essentially be viewed as a sequential polymerisation of amino acids proceeding from one end of an mRNA template to the other and therefore, Simha and Zimmerman introduced statistical mechanics based polymeric models for the same \cite{simha1963polymerization, zimmerman1965kinetics}. MacDonald and Gibbs, on the other hand, proposed a lattice model for the mRNA-ribosome interaction \cite{macdonald1968kinetics, macdonald1969concerning}. This lattice-based probabilistic approach, which later came to be known as the ``Totally Asymmetric Simple Exclusion Process" or ``TASEP", was extended and used in the 1970s and early 1980s to address various problems in regulating protein synthesis and translation \cite{hiernaux1974some, lodish1974model, heinrich1980mathematical, godefroy1981role}. After a brief hiatus until the early 2000s, where the TASEP framework saw most of its application to vehicular traffic flow \cite{schadschneider2000statistical}, the last 20 years has again gained momentum in terms of a renewed interest in the application of TASEP and related models to the translation mechanism \cite{chou2003ribosome, shaw2003totally, lakatos2003totally, chou2004clustered, lakatos2005steady, mehra2006algorithmic, basu2007traffic, dong2007towards, zouridis2008effects, cook2009competition, garai2009fluctuations, garai2009stochastic, romano2009queueing, brackley2010limited, ciandrini2010role, cook2010power, mier2010origins, sharma2010quality, reuveni2011genome, ciandrini2013ribosome, marshall2014ribosome, zhao2014mrna, racle2015noise, gorgoni2016identification, korkmazhan2017dynamics, bonnin2017novel, datta2018influence, park2019inverted, szavits2020dynamics, kavvcivc2021token, margaliot2021variability, sharma2022extrinsic, Jain2022, katz2022translation, jain2022modeling}. 
\\~\\
For developing a basic TASEP model for protein synthesis, the process of translation can be simplified into three main steps, viz. (i) Initiation where the larger ribosomal subunit and the ternary complex come together and incorporate the initiator tRNA on the start codon, (ii) Elongation where the mRNA bound ribosome machinery moves along the mRNA and synthesizes the protein one amino acid at a time and (iii) Termination which is essentially removal of the ribosomal machinery and the complete synthesized protein from the mRNA at the stop codon. The incorporation of these steps are important considerations while formulating the framework for realistic biological systems, such as the translation mechanism in prokaryotic (bacteria \textit{E. coli}) and eukaryotic (yeast \textit{S. cerevisiae}) model organisms. Some of the parameters useful for its implementation are listed in Table \ref{tab:parameters}.
%
%
\begin{table}
\setlength{\tabcolsep}{8pt}
\renewcommand{\arraystretch}{1.5}
\begin{center}
    \begin{tabular}{|p{4.5cm}|p{3.5cm}|p{3.5cm}|}
         \hline
         Feature&\textit{E.coli}&\textit{S. cerevisiae}  \\
         \hline
         Initiation rate \([s^{-1}]\) & \(0.001 - 3.402 \) \cite{gorochowski2019absolute} & \(0.005 - 4.2\) \cite{ciandrini2013ribosome} \\
         \hline
         Elongation rate [codon/\textit{s}] & \(3.9 - 63.7 \) \cite{rudorf2015protein} & \(0.99 - 15.14\) \cite{ciandrini2013ribosome}  \\
         \hline
         Termination rate \([s^{-1}]\) & \(\beta \gg \alpha\) \cite{szavits2020dynamics} & \(\approx 35\) 
        \cite{savelsbergh2003elongation} \\
         \hline
         mRNA length [codons]& \(8-2358\) \cite{10.1093/nar/gkac1052} & \(17-4911\) \cite{10.1093/nar/gkac1052} \\
         \hline
         mRNA half-life \([min]\)  & \(1-49\) \cite{esquerre2014dual} & \(3-300\) 
 \cite{geisberg2014global} \\
         \hline
    \end{tabular}
\end{center}
\caption{Some important parameters while formulating the TASEP based framework for model organisms.}
    \label{tab:parameters}
\end{table}
\\~\\
The parameters listed in Table \ref{tab:parameters} are taken from either purely experimental studies or studies where modeling and further analyses were carried out using data from experiments. Specifically, the initiation rate for E.coli was determined though a combination of sequencing approaches that were then subsequently used to calculate chromosomal ribosome binding site strength in units of ribosomes per second \cite{gorochowski2019absolute}. The initiation rate for \textit{S. cerevisiae}, on the other hand, was calculated in a study that modeled a translation event as a TASEP and matched ribosome density for different initiation rates to experimental data to find optimal initiation rates genome wide \cite{ciandrini2013ribosome}.
The elongation rates of E.coli were inferred from a theoretical frame work of protein synthesis, where codon-specific elongation rates were calculated based on codon usage and tRNA concentrations \cite{rudorf2015protein}. In a study on protein production relative to transcript abundance for the case of \textit{S. cerevisiae} \cite{ciandrini2013ribosome}, the multi step elongation event was simplified into two steps, viz., cognate tRNA capture followed by translocation of the ribosome complex. Here, the translocation rate which is codon independent was taken from an earlier experimental study \cite{savelsbergh2003elongation} and the cognate tRNA capture rates, which are codon dependent were calculated by considering tRNA gene copy number and wobble effect.
The termination rates in the case of prokaryotes are generally larger than the initiation rates \cite{szavits2020dynamics} and in yeast it can be considered similar as that of translocation \cite{arava2003genome}. 
The mRNA length, defined here as the number of codons between the first start codon to the stop codon, was determined by converting protein lengths for both bacteria and yeast from the uniport data base \cite{10.1093/nar/gkac1052}.
The mRNA half-life was determined from an experimental study designed to understand the effects of growth rates (transcription rates) on mRNA stability in E.coli \cite{esquerre2014dual}. The range listed in Table \ref{tab:parameters} corresponds to mRNA half-life for a growth rate of \(0.63 \textrm{  h}^{-1}\). In the case of yeast, the mRNA half-life was determined from a study that sought to quantify mRNA levels at different time points after transcription inhibition \cite{geisberg2014global}.
\\~\\
\begin{figure}[H]
    \centering
    \includegraphics[width=130mm]{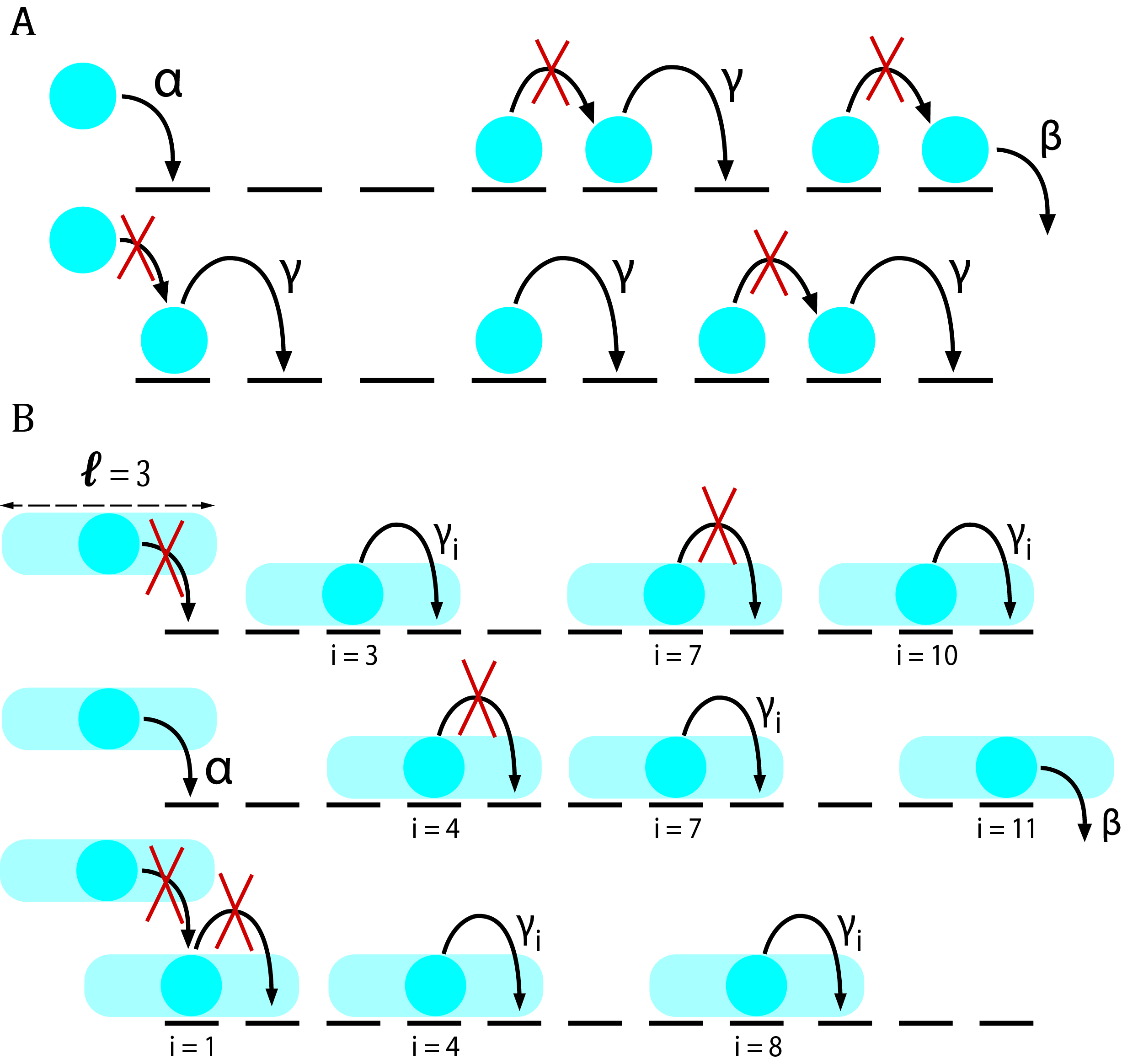}
    \caption{
    (A) Classic TASEP and (B) $\ell$-TASEP. In classic TASEP models of translation $(\ell=1)$ each lattice site corresponds to \(10 - 12\) codons and ribosomes are point particles moving along the lattice strictly following the exclusion principle and provides a ribosome level resolution of the system. In the case of $\ell$-TASEP (represented here with an $(\ell=3)$ particle with active/reaction site at the center)  each lattice site corresponds to a codon and ribosomes are extended particles  physically occupying more than one site on the lattice. In this case, it is necessary to explicitly specify the ribosome-tRNA binding sites to realize an accurate codon level picture of the system.  
    }
    \label{fig:tasep}
\end{figure}
\noindent
Having understood the key steps in the process of translation, one can now model the entire ribosome-mRNA system through the TASEP framework. Disregarding secondary structures of mRNA allows for a simplification of the mRNA template into a linear lattice having L lattice sites. Next, depending on the level of detail required, each lattice site can be considered to correspond to a set of codons for ribosome level resolution (see Fig. \ref{fig:tasep}A) or a single codon for codon level resolution (see Fig. \ref{fig:tasep}B). In the more detailed formalism, the ribosome footprint, which is approximately 10-12 codons in length \cite{skjondal2007dynamic}, is considered explicitly \cite{andreev2018tasep, szavits2020dynamics, jain2022modeling}. However, many other studies that look into density and current of ribosomes consider the ribosomal complex as a point particle \cite{bonnin2017novel, korkmazhan2017dynamics, sharma2022extrinsic} as the results obtained this way are comparable to the system where ribosome footprint is considered explicitly \cite{shaw2003totally}.  
The exclusion principle and total asymmetry of the TASEP model perfectly fit the ribosome movement on the mRNA template. The ribosome binds the mRNA lattice at the 5' end (\(i = 1\)) with a rate \(\alpha\), then jumps to the next lattice site with a rate \(\gamma \) and finally exits the mRNA at the 3' end (\(i = L\)) with a rate \(\beta\). Ribosomes follow the exclusion rule throughout the process, \emph{i.e.,} no two ribosomes can physically occupy the same lattice position at the same time. This process is represented schematically in Fig. \ref{fig:tasep}. For a lattice of length L there can be a maximum of  \( L-1\) elongations, an initiation and a termination, which constitute a maximum reaction pool of size \(L+1\). Depending on the state of the lattice, possible reactions are picked from this reaction pool by strictly following the exclusion rule and are used for calculating the total propensity. The classical TASEP model which can be used for ribosome level resolution of translation follows exclusion rules as explained previously but the $\ell$-TASEP (with $\ell = 10$) which can be used for codon level resolution requires a different set of rules. The algorithms for both these TASEP frameworks are represented as a flow chart in Fig. \ref{fig:flowchart_tasep}. In the more detailed framework (codon level resolution), the ribosome physically occupies \(10\) codons ( 5 sites to the left and 4 sites to the right\cite{ingolia2009genome} ) while decoding a single codon at the A site. During initiation the ribosomal subunits come together and form the ribosomal complex with the initiator tRNA over the P site, which is located at the $5^{th}$ lattice site occupied by the ribosome. For initiation to occur, the first $11$ lattice sites need to be vacant. This is because the start codon of the mRNA is positioned at the P site of the initiating ribosome and therefore the next initiation can only happen once the first elongating ribosome is at the 12th codon. Physical distance between two ribosomes actively decoding the codons during the elongation step needs to be greater than the length of the extended object ($\ell$), in this case 10 and for termination, the A site of the ribosome needs to be over the stop codon (\emph{i.e.,} $i=L$). Additional features such as capture of cognate tRNA by ribosome, availability of factors involved in each step, etc can be included depending on the level of detail required for the study of the system.
\\~\\
\noindent
The current lattice can be updated sequentially or in parallel \cite{rajewsky1998asymmetric} using the Gillespie algorithm \cite{gillespie1977exact, gillespie2007stochastic} after exclusion rules have been checked. Sequential update can be done in one of two ways, viz. (i) in order, where updation starts at one end of the lattice site and sequentially moves to the other, for example, start at the right end, check exclusion and update $i=L$ and $i=L-1$ then $i=L-1$ and $i=L-2$ and so on and when the left end $i=1$ is updated allow particle injection; or (ii) at random, where a site or reaction is chosen at random and updation for a particular lattice site is carried out in a small time step. For parallel update, on the other hand, the exclusion rule is checked for all the lattice sites at once and updated. Therefore, for a lattice whose length is even, this can be done in pairs. In both kinds of update and for both the TASEP models, whether classical or detailed, ribosomes follow the exclusion rule and unidirectional movement at all times. The protein production rate can then be identified as the ribosomal current on the mRNA lattice. 
\\~\\
Particle movement along a lattice is an interesting non-equilibrium problem in itself and some of the early works focused on looking at the steady and non-steady states of this system \cite{parmeggiani2003phase, lakatos2005steady}. In the context of mRNA translation then, this turns into looking at ribosome dynamics on a newly transcribed mRNA and an mRNA with ribosome density close to the steady state level \cite{szavits2020dynamics}. In particular, this TASEP based model was developed to mimic an experimental study where translation dynamics was probed in real time by tagging ribosomes with green fluorescent protein \cite{yan2016dynamics}. This model made use of realistic kinetic parameters in the stochastic simulations to then make predictions about ribosome dynamics before and after reaching steady state and ribosome runoff dynamics after initiation was blocked. Further, since the translation process involves many different components, such as ribosomes, tRNA and mRNA itself, competition for resources is another problem that is of interest. The earliest model that looked into competition with respect to protein synthesis considered competition between mRNAs for limiting components of the initiation apparatus \cite{godefroy1981role}. Specifically, the model considered initiation as a multi-step process in which the initiation site needs to be activated by binding to a discriminatory factor before undergoing recognition by the 40S subunit. In a system with saturated amounts of elongation and termination factors, translation of mRNAs were shown to be affected by their affinity to the discriminatory factor. The model, therefore, showcased how a single step could also have an impact on competition and suggested that the presence of multiple such steps could be the reason for variations in the kinetics of cellular processes between cell lines and cells in different environmental conditions. Modeling studies also looked into tRNA competition since a pool of different tRNAs can, in practice, bind to the same set of codons. The kinetics of protein synthesis and translation in this context was tackled both through Michelis-Menten based \cite{zouridis2008effects} and TASEP based \cite{brackley2010limited, gorgoni2016identification} deterministic and stochastic models, respectively. In fact, the TASEP model of codon resolution with a particle size of 9 and two step elongation (\emph{i.e.,} recognition of cognate tRNA and translocation to next codon) was used to model translation on 5500 yeast open reading frames (ORFs) with ORF specific initiation and constant termination \cite{gorgoni2016identification}. The simulation carried out in an environment of invariant ribosome concentration and charged tRNA was able to predict mRNA targets \sout{species} that were sensitive to single tRNA species concentration level, that were then confirmed experimentally. The study was able to show that low tRNA concentrations can cause ribosome queuing, which in turn can be overcome by controlling initiation rates. This proposed the presence of initiation and elongation regulated mRNA species in the cell. Similarly, mRNA translation under the condition where the pool of ribosomes is limited in comparison to the number of mRNAs causes the total protein production to saturate as more mRNAs are added. This phenomenon was predicted theoretically through an RFM framework, as described previously \cite{katz2022translation}. Prior to this, another discrete state Markov modelling based whole cell simulation study carried out with experimentally determined numbers of mRNAs, tRNAs and ribosomes was not only used to predict the initiation rates of endogenous genes of \textit{S. cerevisiae}, but could also elucidate the parameter regimes that led to increased protein production. Protein production was also found to be limited by ribosome availability, which could nevertheless be overcome by regulating initiation and elongation rates \cite{shah2013rate}. Another study looked at a similar system with the distinction that the mRNAs could be of varying lengths \cite{cook2009competition}. In this, multiple TASEPs of different lengths were coupled to a finite reservoir, such that the initiation rates now depended on the number of particles (\(N_{p} \)) in the pool. The effective initiation rate was then defined as \(\alpha_{eff} = \alpha f(N_p)\), 
 with $f(N_p)$ varying between 1 and 0 for the two extremes where ribosomes are abundant and exhausted, respectively. This model was used to investigate the system in the LD, HD, MC and SP regimes which helped understand delocalization of the domain wall as a result of competition by other TASEPs.
 \\~\\
 TASEP based modeling studies have also been used to predict various phenomena in cellular systems. For example, it is known that certain upstream open reading frames (uORFs) present in mRNA control translation and render these mRNAs resistant to stress \cite{andreev2015translation}. A plausible explanation for this was provided by a simulation study of a detailed TASEP model of codon resolution that had two types of particles, \textit{i.e.,} scanning and elongating ribosomes, each occupying 10 codons \cite{andreev2018tasep}. In this model system, all the particles are injected into the lattice as scanning ribosomes representing ribosome binding at the ribosome binding site (RBS) at the 5' UTR. The scanning ribosomes then either transform to the elongating ribosomes at the start codon of uORF with a high probability or move downstream representing leaky initiation. The leaked scanning ribosomes in turn drop off from the lattice when upstream elongating ribosomes collide with them. This model then helped with understanding the role of ribosome loading rate, pausing and slowing of elongation, strength of the uORF start codon and reinitiation of uORF ribosomes. Ribosome collisions and dropoff were also explicitly modeled in other studies to look into how they affect ribosome density on the mRNA \cite{bonnin2017novel, scott2019power, park2019inverted, jain2022modeling}. 
 \\~\\
 Other questions concerned ribosome density and traffic in the presence of slow codons which caused traffic jams and led to a change in the translation kinetics \cite{chou2004clustered, korkmazhan2017dynamics}. A TASEP based model was considered where an mRNA was divided into a signaling and a non-signaling region and depending on the ribosome occupancy in the two regions assigned it as freely diffusing or membrane-bound, respectively. This was then used to look into the spatial localization of the translated proteins on the membrane and how the subsequent spatial heterogeneity and distribution of protein cluster sizes varied with different initiation rates and the presence of slow codons \cite{korkmazhan2017dynamics}. Yet other studies went into further details with genome wide analysis of translation where the initiation, elongation and termination rates varied as per the corresponding sequence in the codon \cite{mehra2006algorithmic, reuveni2011genome, ciandrini2013ribosome} In particular, experimentally obtained ribosome densities were combined with TASEP to discover that codon arrangement has an essential role to play in translation dynamics \cite{ciandrini2013ribosome}. Here, the TASEP model considered was again of codon resolution with extended particles of size 9 and a two step elongation process. This was then used to estimate mRNA specific in vivo translation initiation rates. The model predicted ribosome density and current as a function of initiation rates, which were then compared with experimental values to find physiological initiation rates in Yeast. The study was also able to look at codon usage effects, responsiveness of translation to change in initiation rates and helped in identifying functional groups of proteins most susceptible to translation control through initiation. Such theoretical and simulation studies have also been extended to include the effects of noise \cite{datta2018influence, sharma2022extrinsic}, tRNA and ribosome distribution \cite{margaliot2021variability, racle2015noise}.
\\~\\
In addition to all the above mentioned features, depending on the aspect of protein synthesis under study, increasing complexity in reactions or dynamics can be incorporated. Such modeling studies have been developed to account for the kinetics and dynamics of translation in the context of different sizes of ribosomes \cite{lakatos2003totally}, ribosome recycling \cite{marshall2014ribosome}, finite resources of ribosomes and tRNAs \cite{kavvcivc2021token} and circular mRNA \cite{chou2003ribosome,raveh2015ribosome}, among others.


\section{TASEP based modeling of other biological processes}
\noindent
TASEP eventually became a standard mathematical modeling framework \cite{zia2011modeling}, not only for translation, but also for other systems with unidirectional flow of particles along a lattice, such as transcription, intra-cellular transport and fungal hyphal growth.
\\~\\
\noindent
\textbf{TASEP for modeling transcription.} Transcription is a process similar to translation in which a DNA segment (or, gene) is stochastically elongated through specific enzymes called RNA polymerases (RNAPs). An RNAP binds to the promoter region upstream of the gene and moves forward. Once it recognizes the gene, it sequentially and successively adds nucleotides to the growing RNA chain until it encounters a stop signal. At this point, the complete RNA is released. Therefore, in essence, the process of transcription is also a problem that can be identified with unidirectional particle hopping on a lattice. This unidirectional movement along the lattice is often stalled and the resulting transcription kinetics that gets affected due to these pauses is a problem of interest. In this context, random pauses \cite{Klumpp2008, Klumpp2009, wang2014minimal}, 
backtracking restart after stalling \cite{li2015theoretical, zuo2022density} and RNAP stalling due to nucleosome binding and unbinding kinetics on the DNA that lead to transcription bursts and traffic jams \cite{mines2022slow}, all affect the overall kinetics of transcription and have been studied. As is the case for slow nucleosome dynamics leading to polymerase pausing \cite{mines2022slow}, many models have focussed on the binding and unbinding of defects to the DNA lattice sites that then cause unexpected changes to transcription kinetics and polymerase dynamics \cite{waclaw2019totally, szavits2020current}. These modeling frameworks are generally referred to as dynamic disorder or defect disorder TASEP models. Specifically, a TASEP model with dynamic disorder is one in which a blockage or a defect appears and disappears stochastically across the lattice and when present, slows down or completely stops the polymerase from moving along the chain \cite{waclaw2019totally}. Further analysis of such a model shows that the current-density relation assumes a quasiparabolic form which is similar to the regular TASEP framework without any defect related pausing \cite{szavits2020current}. An extension of the ordinary TASEP framework was also developed and analyzed in order to understand transcription during DNA replication. In the presence of a replication fork, the RNAP engaged in transcription can suffer either co-directional (due to higher speed of replication than transcription) or head-on encounter with the replication fork (movement of  replication fork in both directions from the Origin) which can again cause stalling of DNA replication. A TASEP based two species exclusion model was able to show how transition between density phases affects not only the replication time but also the percentage of successful replication events. \cite{ghosh2018biologically}.
\\~\\
\textbf{TASEP models for intracellular transport.}
Intra cellular transport by molecular motors are key to diverse cellular activities like movement, division and transport. Kinesins and dyneins are motor proteins that use ATP-derived energy to transport a variety of intracellular cargoes. Kinesins carry molecules toward the plus-ends of the microtubules, where beta-tubulin is exposed and dyneins carry them towards the minus-ends, where alpha tubulin is exposed \cite{gennerich2009walking}. TASEP models have also been used to study this system of molecular transport. The microtuble filament, that acts as a platform for this transport, is considered as a one dimensional lattice composed of equally spaced binding sites for molecular motors. Further, sterical hindrance ensures that each site can only bind to a single enzyme. Additionally, the binding rate at the two ends are distinct from each other as they depend on the enzyme reservoir densities. All these can be used as system conditions that can then undergo dynamics as per the TASEP framework \cite{frey2004collective}. Such a model was used to study the formation of a morphogen gradient along the cytoskeleton \cite{bressloff2018bidirectional}. In addition, transport based TASEP models were also used to study the collective dynamics \cite{klumpp2003traffic, klein2005filament, ciandrini2014stepping, ciandrini2014motor, kuan2016motor, rank2018crowding} and detachment of molecular motors under the conditions of crowding on the microtubule \cite{rank2018crowding}. Analysis of various statistics of the motor dynamics suggested spatial correlations between co-moving clusters and the intermittent transition of the motor to an inactive state.
\\~\\
%
%
%
%
\textbf{TASEP modeling for fungal hyphal growth.}
Hyphae is a long, filamentous structure in a fungus and is the main mode of vegetative growth in it. The hyphae growth is facilitated by continuous movement of materials into its tip region. The tip extension progresses rapidly towards nutrient rich areas that eventually leads to the penetration of cell walls or insect cuticles. This movement of molecules from bulk to the tip of the fungal body can be simplified as the dynamical movement of mass along a quasi-one-dimensional hyphae with particles hopping on the growing lattice while following the mutual exclusion principle \cite{sugden2007dynamically, evans2007exclusion}. From phase transition studies of this model, the existence of growing hyphal system in the low density phase was observed and was speculated to be necessary for the branching phenomena.


\section{Discussion}
\noindent
We have, in this review, provided an overview of the TASEP framework and some of its key features. In addition, we have also provided several examples that show how useful this framework of ``particle hopping on a lattice" is and how it can be applied to a varied set of systems. This framework that was first written in terms of the driven lattice gas model and later as ASEP and TASEP, has seen tremendous development over the last 40 years. Around the same time that the driven lattice gas model was proposed \cite{van1985excess}, another model that looked into stochastic growth of interface heights was also developed by Kardar, Parisi and Zhang \cite{kardar1986dynamic}. The Kardar-Parisi-Zhang (KPZ) equation, as proposed in their seminal paper is a stochastic partial differential equation that gives a distribution of growth heights and was later realized to be of importance for explaining a variety of stochastic growth systems \cite{corwin2012kardar}. It was also realized that this equation and properties associated with it created a new universality class, known as the KPZ universality class. Further, it was worked out that the ASEP framework could essentially be described via the KPZ equation and could, therefore, be interpreted in terms of growing interfaces \cite{derrida1998exactly}. Exact formulas for transition probabilities for N particles in the ASEP framework thus puts all the simple exclusion processes in the KPZ universality class \cite{corwin2012kardar}. Exploring TASEP within this universality class has therefore been a topic of interest in statistical physics. 
%
\\~\\
Further, the simple TASEP framework which considered the particle to occupy just one lattice site was later extended to a system where the particle could, in fact, span a physical space of more than one lattice site. This was a more realistic model of ribosome occupancy on an mRNA lattice and was termed TASEP with extended objects \cite{shaw2003totally, garai2009fluctuations, brackley2010limited}. The ribosome flow model, as discussed earlier, is another extension which considers mean-field solutions of the TASEP framework \cite{reuveni2011genome, zarai2017optimal, zarai2017ribosome, nanikashvili2019networks, Jain2022}. A mean field theory of TASEP was also recently proposed to look at two-species particle hopping with distinct entry and exit rates \cite{bonnin2022two}. Another formalism recently applied to the same problem of translation used a random matrix theory based approach to incorporate a random distribution of kinetic rates into the standard TASEP model \cite{margaliot2021variability}. 
\\~\\
%
Even with the above elaborated continued development of the TASEP framework in the context of biological movement and transport at the intracellular level, it is important to note that it cannot be seen as a ``one size fits all" kind of theory. The TASEP model is, in essence, a simplified lattice framework and therefore ignores many of the intricate details of the participating players. For example, it cannot be used to include even coarse-grained details of the structure of the individual ribosomes, tRNAs and even the mRNAs themselves. Second, the formalism in TASEP can only be developed, at most, in terms of the kinetics of the individual binding-unbinding events and not in terms of the intricate binding sites. Further, the temporal and spatial dynamics of the ribosomes can only be seen on and within the vicinity of the mRNA lattice. However, despite the simplicity of the framework, it can still be used to explain many of the recent experimental observations related to the kinetics of translation and transcription.   
\\~\\
Extensions of the early TASEP framework that incorporate further details of the real system have, in fact, come about as new and more sophisticated experiments are revealing more about transcription and translation. Thanks to recent technological advancements, kinetics data can now be obtained at single molecule resolution \cite{wu2016translation, morisaki2018quantifying, khuperkar2020quantification} in living cells and several detailed mechanisms are therefore being charted out. A recent study has identified and segregated the contribution of RNA degradation to intrinsic and extrinsic noise \cite{Baudrimont2019}. Experiments performed to understand translation have revealed that the initiation process is controlled by RNA folding and ribosome binding kinetics \cite{espah2016translation}. Other aspects of translation have also been studied experimentally. For example, single tRNAs have been tracked to look at the effect of their kinetics on overall translation \cite{volkov2019tracking}, ribosome dropoff has been quantified in \textit{E. coli} bacteria \cite{sin2016quantitative} and temporal fluorescence signals have revealed codon-specific protein synthesis rates \cite{haase2018decomposition}. In addition to the kinetics of translation and its breakdown into its constituent processes carried out by individual components, spatial dynamics is another aspect that is of interest. Single molecule experiments have shown that the association of mRNAs with the endoplasmic reticulum is dependent on the translation mechanism \cite{voigt2017single} . Another imaging study has quantified the spatial heterogeneity of mRNAs in \textit{Drosophila} embryos and looked at the role of clustered mRNAs in translation efficiency \cite{dufourt2021imaging}. 
\\~\\
Apart from the above mentioned single molecule experimental studies, large scale transcriptome data, which give an insight into the mRNA levels of individual transcripts, have also contributed to the understanding of gene expression, gene control and the identification of novel transcripts \cite{nagalakshmi2008transcriptional}. However, mRNA levels are poor proxies for protein production as they are still subject to extensive regulation by cellular and environmental conditions \cite{baek2008impact}. This is where ribosome profiling, which is a combination of polysomal profiling and RNA sequencing technology, has an advantage as it allows for the measurement of gene expression at the translational level \cite{gobet2017ribosome, calviello2017beyond}. A translating ribosome traverses about 30 nucleotides on the mRNA and can protect the fragment from RNAse digestion. The traversing ribosomes can be treated with chemicals such as cycloheximide to stabilize and fix them on the mRNA and then when treated with endonuclease, such as  micrococcal nuclease, can lead to the degradation of unprotected regions while conserving ribosome-protected regions. These protected RNA fragments corresponding to the typical footprint of ribosomes can be purified, sequenced and aligned to the reference genome to understand actively translated mRNA regions. In this context,a recent ribosome profiling method has revealed pauses in translation at the single-codon resolution \cite{mohammad2019systematically, vaninsberghe2021single}. These high precision measurements of the in-vivo translation process allow for a better understanding of transcription and translation, leading to the development of more accurate models of protein production  \cite{calviello2017beyond,erhard2018improved}. 
\\~\\
To this end, TASEP models of translation, working with ribosome hopping as the functional process are directly benefiting from newly emerging ribo-seq data \cite{zur2016predictive, neelagandan2020determines, yadav2021quantitative}. Over the years, several methods have been developed for the extraction of ribosome density and other kinetic rates from ribosome profiling data. Additionally, there has also been an advancement in the understanding of various properties of ribosome kinetics and the translation process as a whole. These include use of the change point algorithm on the data for determining ribosome density \cite{zupanic2014detecting}, analysis of ribosome density obtained from the data to determine translation rate \cite{gritsenko2015unbiased} and the use of chemical kinetic principles for extracting translation kinetic rates of individual processes \cite{sharma2019chemical}. These methodologies, therefore, have also helped in exploring correlations between tRNA concentration and translation kinetics \cite{dana2014effect}, in the understanding that codon translation rates are randomly distributed leading to nearly equal average translation rates across transcripts  \cite{sharma2018determinants}, uncovering the fact that ribosome queuing is prevalent and therefore leads to frequent traffic jams \cite{diament2018extent} and revealing the variation in translation elongation rates and how that is correlated to the translation initiation rates \cite{riba2019protein}. However, even though the ribosome profiling data can provide the relative occupancy and density of ribosomes on the mRNA at single codon resolution, the translation rates themselves may depend on several other molecular factors such as the tRNA abundance, context dependent variation of codon translation rates and the mRNA structure itself. Therefore, apart from the riboseq data, one must also take into account the particular kind of mRNA and transcript being translated for a complete understanding of the variation in translation rates. These could then help in reliable prediction and understanding of post-translational effects as well.
\\~\\
%
All these single molecule and ribosome profiling experiments reveal many details of the mechanism of translation and temporal and spatial dynamics of the individual components of the translation process. These experiments, therefore, open up many avenues for development of more detailed and robust theories that can explain and predict existing phenomena in living cells. With the success that the TASEP framework has already achieved, it seems to be well primed to be extended to explain the recent experimental observations. In fact, modifications and extensions of the classical TASEP with the purpose of including intricate particle dynamics gives rise to models that are better representations of biological processes in cellular environments. These detailed models can be perturbed to understand the impact of each step of the process or individual particle properties in the pathway. In addition to the experimental observations, one of the most sought after properties of TASEP with open boundaries is the existence of phase transitions, even in a one dimensional lattice. These phase transitions can provide insights into particle queuing and congestion observed in biological transport and translocation systems and, therefore, into cellular control at the microscopic level. Understanding where in the TASEP phase diagram do we find natural systems and examining conditions for phase transitions in biological models can shed light on the efficiency and robustness of biological processes. These and many other open biological problems are yet to be studied theoretically. 
\\~\\
Computational and mathematical models of translation that closely represent the real system can also be used to study and predict the role of factors and processes by carrying out specific modifications. Even though translation is an evolutionarily conserved and tightly regulated process, dysfunctions of various steps/processes involved can lead to diseases. For example, alterations in the structure and function of ribosomal components cause a class of diseases called ribosomopathies, which includes Diamond-Blackfan anemia, 5q-syndrome and Schwachman-Diamond syndrome \cite{nakhoul2014ribosomopathies}. Mutations in non-ribosomal proteins and translational factors and defects in amino-acyl-tRNA synthetase can impact localization of charged tRNAs and therefore the accuracy of protein translation \cite{scheper2007translation}. Further, dysfunction of  translation of the mRNA encoded in the mitochondrial DNA are known to be a cause for mitochondrial diseases \cite{gorman2016mitochondrial}. Therefore, TASEP models which can adapt to any level of resolution required for the system under consideration, can be used to shed light on the underlying causes and progression of these diseases. Incorporation of cognate tRNA diffusion and translocation as separate events, for example, can equip the model with the ability to probe mitochondrial diseases caused by tRNA modifications. The structural changes caused by mutations in ribosomal proteins can be incorporated by considering the probability of binding a defective or a normal ribosome to the lattice during a translation initiation event. The modelling framework, therefore, still has a lot of scope for exploration towards the understanding of these diseases.
\\~\\
In conclusion then, TASEP models have provided theoreticians with a powerful and promising framework that can be used to understand several different biological processes. Therefore, as things stand, it appears that the framework will continue to be exploited and explored for years to come.


\section*{Acknowledgments}
This work is supported by the Science and Engineering Research Board (SERB) MATRICS Research Grant (Ref. No. MTR/2022/000655) awarded by the Department of Science and Technology (DST), India.

\bibliographystyle{unsrt}
\bibliography{refs.bib}

\end{document}